\documentclass[twocolumn,showpacs,amsmath,amssymb]{revtex4}

\begin{document}

\topmargin -2cm
\title{Stationary Rotating Black Hole Exact Solution within Einstein--Nonlinear Electrodynamics}
\author{Alberto A. Garc\'\i a--D\'\i az}
\altaffiliation{aagarcia@fis.cinvestav.mx}
\affiliation{Departamento~de~F\'{\i}sica.
\\~Centro~de~Investigaci\'on~y~de~Estudios~Avanzados~del~IPN.\\
Apdo. Postal 14-740, 07000 M\'exico DF, MEXICO.\\}

%\Large
\date{\today}

\begin{abstract}
The first exact rotating charged black hole solution to the
Einstein--nonlinear electrodynamics theory is reported. It is
equipped with mass, rotation parameter, electric and magnetic
charges, and three parameters due to the electrodynamics: $\beta$ is
associated to the potential vectors $A_\mu$ and ${}^{\star}P_\nu$,
and two constants, $F_0$ and $G_0$, related to the presence of the
invariants $F$ and $G$ in the Lagrangian $L(F,G)$. This Petrov type
D solution is characterized by the Weyl tensor eigenvalue $\Psi_2$,
the Ricci tensor eigenvalue $S=2\Phi_{(11)}$, and the scalar
curvature $R$; it allows for event horizons, exhibits a ring
singularity and fulfils the energy conditions. Its Maxwell limit is
the Kerr--Newman black hole.
\end{abstract}

%$FILE:C:PlebCarterNL2021.tex$
\vspace{0.5cm}

\maketitle
%\tableofcontents

 Two decades ago we reported the first exact regular
spherical symmetric back hole solution \cite {AyonGarcia:1998}, in
the framework  of Einstein--NLE equations opening a fruitful and
active  area of research in the search for {\bf exact solutions}.
Previous to this work, there was known a {\it model}--the Bardeen
model \cite {Bardeen68}--whose metric determines an Einstein tensor,
which, via the Einstein equations ``{\it defines}'' a matter tensor
$T_{ab}$ fulfilling the energy conditions, and therefore, one could
associate to it a viable energy matter-field tensor, although in the
Bardeen publication, no mention to a possible matter content was
mentioned. Only later \cite{AyonGarciaOnBardeen}, we succeeded to
derive a nonlinear magnetic electrodynamics source for the Bardeen
model, since then it acquires the status of exact solution in the
NLE frame.

Recently, for static spherical symmetric metrics the general exact
solution coupled to NLE \cite{Garcia-Diaz:2019acq}, with an
arbitrary structural metric function, which, via a pair of
independent Einstein equations, allows one to derive the single
associated field tensor component $\mathcal{E}$, and the
Lagrangian--Hamiltonian field function $\mathcal{L}$--$\mathcal{H}$,
which determine the entire solution, should it be singular or
regular.

Referring  to  exact solutions we adopt the criteria of Hawking an
Ellis,\cite {Hawking73}, pag 117:``We shall mean by an {\it exact
solution} of the Einstein's equations, a space--time
$(\mathcal{M},g)$ in which the field equations are satisfied with
$T_{ab}$ the energy momentum tensor of some specified form of matter
which obeys postulate (a) (`local causality') of chapter 3 and some
of the energy conditions of \S 4.3\ldots''  In this respect we
consider as {\bf solutions} those fulfilling the HE criteria of
exact solution, and reserve the name of {\bf models} for those
results derived by the ``metric--defined matter tensor'' procedure
fulfilling reasonable energy conditions. Therefore in this sense,
all the reported until now Kerr--like \cite{TorresFayos17} rotating
black holes are {\bf models}, they would become solutions, if
someone should be able to determine  the matter-field
energy--momentum tensor $T_{ab}.$

This first exact solution describes a stationary rotating black hole
endowed with several parameters; it fulfils a set of four
generalized ``Maxwell equations'' for the electrodynamics fields
$F_{\mu\nu}$ and $P_{\mu\nu}$ and two independent Einstein--NLE
equations related with the two independent eigenvalues of the NLE
energy--momentum tensor. The NLE is determined by a Lagrangian
function $L$ constructed on the two electromagnetic invariants $F$
and $G$, $L(F,G)$.

To avoid misinterpretations due to the use of different definitions
and sign conventions by different research groups, we give a
self--contained resume of the NLE we are dealing with.\\
\noindent It is well known the standard definition of the energy
momentum tensor in general relativity: following
Stephani,\cite{StephaniBook1990}, \S 9.4. from the variational
principle upon the action
\begin{eqnarray}\label{actio}
W=\int{\sqrt{-g} \,d^4x\,\left(R/2 +\kappa L_M \right)},
\end{eqnarray}
taking into account $ \delta \sqrt{-g}=\delta
g^{\mu\nu}\frac{\partial}{\partial
g_{\mu\nu}}\sqrt{-g}=\frac{1}{2}\sqrt{-g}g^{\mu\nu}\,\delta
g_{\mu\nu}$, the variation yields
\begin{eqnarray}
\delta W=-\frac{1}{2}\int{(R^{\mu\nu}-\frac{1}{2}R\,g^{\mu\nu}
-\kappa\,T^{\mu\nu})\,\sqrt{-g} \delta g_{\mu\nu}}\,d^4x.
\end{eqnarray}
Hence from the vanishing of this variation one arrives at
 the Einstein equations in the presence of matter and fields described by an
energy--momentum tensor $T^{\mu\nu}$
\begin{equation} R^{\mu\nu}-\frac{1}{2}R\,g^{\mu\nu}=\kappa
T^{\mu\nu};\,
T^{\mu\nu}=\,\frac{2}{\sqrt{-g}}\frac{\delta(\sqrt{-g}\,L_M)}{\delta\,g_{\mu\nu}},
\end{equation}
where $L_M$ stands for the matter--field Lagrangian $L_M$,

Again from \cite{Hawking73}, \S 2.8: `` One can use the existence of
a metric to define (in a given basis) the form
${\bf{\eta}}=|g|^{1/2}{\bf {\epsilon}}$, where $g=\text
{det}(g_{ab})$.This form has components $\eta_{ab\ldots d}=n!
|g|^{1/2}\,\delta^1_{[a }\delta^2_b\ldots\delta^n_{m]}$. The
contravariant antisymmetric tensor $\eta^{ab\ldots
d}=g^{ar}g^{bs}\ldots g^{dm}$ $\eta_{rs\ldots d}$ has components
$\eta^{ab\ldots cd}=(-1)^{(n-s)/2}\,n!\,
|g|^{-1/2}\,\delta^{[a}_1\delta^b_2\ldots
 \delta^{d]}_n$,
where $s$ is the signature of ${\bf g}$''. Therefore, for a Lorentz
metric, $$\eta^{abcd}=-\,4!\, |g|^{(-1/2)}\,\delta^{[a}_1\delta^b_2
\delta^c_3 \delta^{d]}_4$$ in agreement with the Lichnerowicz
definitions \cite{LichBook1955}:
\begin{eqnarray}\label{Lich}&&
\eta_{\alpha\beta\gamma\delta}=\sqrt{-g}\epsilon_{\alpha\beta\gamma\delta},\,
\eta^{\alpha\beta\gamma\delta}=-\frac{1}{\sqrt{-g}}\epsilon^{\alpha\beta\gamma\delta}.
\end{eqnarray}
We shall adopt these conventions in the definitions of dual tensors.
In the above paragraph we are using quotation marks although our
transcriptions are partial with minor modifications.

%\begin{eqnarray}
%\end{eqnarray}

The electrodynamics is described by an antisymmetric electromagnetic
field tensor $F_{\mu \nu}=A_{\nu ,\mu}-A_{\mu ,\nu}$, together with
its dual field tensor
\begin{eqnarray}{}^{\star}F_{\alpha\beta}=\frac{1}{2}\eta_{\alpha\beta\mu\nu}F^{\mu\nu}
=\frac{1}{2}\sqrt{-g}\epsilon_{\alpha\beta\mu\nu}F^{\mu\nu},\nonumber\\
{}^{\star}F^{\alpha\beta}=\frac{1}{2}\eta^{\alpha\beta\mu\nu}F_{\mu\nu}
=-\frac{1}{2\sqrt{-g}}\epsilon^{\alpha\beta\mu\nu}F_{\mu\nu}
\end{eqnarray}
where the numerical $\epsilon$--tensor is associated to the
4--Kronecker tensor. The nonlinear electrodynamics is constructed
from a Lagrangian function $L=L(F,\,G)$ depending on the
electromagnetic invariant $F$ and pseudo--scalar $G$:
\begin{equation}
F=\frac{1}{4}F_{\mu\nu}F^{\mu\nu},\,{{G}}=\frac{1}{4}{}^\star{{F}}_{\mu\nu}F^{\mu\nu},
{}^\star{{F}}_{\mu\sigma}F^{\nu\sigma} ={{G}}\,\delta^\nu_\mu.
\end{equation}
To construct the energy--momentum tensor $T^{\mu\nu}$ one
accomplishes the variation of $L_M$  with respect to $g_{\mu\nu}$.
In NLE one uses for the Maxwell limit the Lagrangian function
$L_{Max}(NLE; F)=F^{\alpha\beta}F_{\alpha\beta}/4$, instead of the
standard Maxwell Lagrangian function $L_{Max}( F)=-F$, thus, to
obtain the Maxwell limit, one has to use the Lagrangian function
$L_M=-L(NLE;F,G)$. Therefore, accomplishing the variations, one
arrives at
\begin{eqnarray}\label{EnegryT1}
-T^{\mu\nu}&&=L\, g^{\mu\nu}-L_F\,F^{\mu\sigma}
{F^{\nu}}_{\sigma}-L_G{F^{\mu\sigma}}{}^\star{F^{\nu}}_{\sigma}\nonumber\\&&=:L\,
g^{\mu\nu}-F^{\mu\sigma}{P^\nu}_{\sigma}.\end{eqnarray} One  always
may introduce in nonlinear electrodynamics a new field
\begin{eqnarray}
P_{\mu\nu}=2\frac{\partial L}{\partial F^{\mu\nu}}=L_F
F_{\mu\nu}+L_{{G}} {}^{\star}{F}_{\mu\nu},
\end{eqnarray}
which one identifies as the $P_{\mu\nu}$ field tensor of Pleba\'nski
\cite {SalazarGarciaPleb1},  or the $p^{kl}$--field tensor of
Born--Infeld.

From the antisymmetric tensor field tensor $P_{\mu\nu}$ one
constructs its dual field tensor ${}^\star P_{\mu\nu}$ and the
invariants
\begin{eqnarray}&&{}^{\star}P_{\mu \nu}:=\frac{1}{2} \sqrt{-g}\epsilon_{\mu\nu\alpha\beta}P^{\alpha
\beta},\, \,{{}^\star{P^{\alpha\beta}}}=-\frac{1}{2\sqrt{-g}}
\epsilon^{\alpha\beta\mu\nu}P_{\mu\nu}, \nonumber\\&&
P=\frac{1}{4}P_{\mu\nu}P^{\mu\nu},\,{{Q}}=\frac{1}{4}
{{}^{\star}{P}}_{\mu\nu}P^{\mu\nu}.
 \end{eqnarray}
The Hamiltonian function $ H(P,Q)$, associated with the Lagrangian
function $L(F,G)$, can be determined by a Legandre transformation
\begin{eqnarray}&&
L(F,G)=\frac{1}{2}F_{\mu\nu}P^{\mu\nu}-H(P,Q); \nonumber\\&&
P_{\mu\nu}=2\frac{\partial L}{\partial F^{\mu\nu}}=L_F
F_{\mu\nu}+L_G\,{{}^\star F}_{\mu\nu},\nonumber\\&&
F_{\mu\nu}=2\frac{\partial H}{\partial P^{\mu\nu}}=H_P
P_{\mu\nu}+H_{{Q}} {}^{\star}{P}_{\mu\nu}.
\end{eqnarray}
The electrodynamics is determined through  the ``Faraday--Maxwell''
electromagnetic field equations, which in vacuo are
\begin{eqnarray}
{{}^\star{F^{\mu\nu}}}_{;\nu}=0 \rightarrow {(\sqrt{-g}{}^\star {F^{\mu\nu}})_{,\nu}=0 },\nonumber\\
{P^{\mu\nu}}_{;\nu}=0
\rightarrow{\left[{\sqrt{-g}L_F\,{F^{\mu\nu}}+\sqrt{-g}L_G{{}^\star
{F^{\mu\nu}}}}\right]_{,\nu}=0},
\end{eqnarray}
which can be written by means of a closed 2--form $d\omega=0$,
\begin{eqnarray*}
\omega=\frac{1}{2}\left(F_{\mu\nu}+{{}^\star
P_{\mu\nu}}\right)dx^\mu\,\wedge
dx^\nu=\frac{1}{2}\left(F_{ab}+{{}^\star P_{ab}}\right)e^a
\wedge\,e^b,\
\end{eqnarray*}
since $F_{\mu\nu}$ and ${{}^\star P_{\mu\nu}}$ are curls.

In nonlinear electrodynamics, the energy--momentum tensor
${T^\mu}_{\nu}$, (\ref{EnegryT1}), allows for two different pairs of
eigenvalues $\{\lambda,\lambda,\Lambda^\prime,\Lambda^\prime\}$. One
can show that a similar property,~i.e., two pairs of different
eigenvalues, is sheared by the field tensors $F_{\mu\nu}$ and
$P_{\mu\nu}$, although their eigenvalues are different.
${T^\mu}_{\nu}$ possesses a non zero trace
\begin{eqnarray}&&\label{EnergyTrace}
-T:=-{T^\mu}_{\mu}=4\,L -4\,L_F\,{F}-4\,L_G\,G.
\end{eqnarray}
On the other hand, taking into account the relation
${F}^{\mu\sigma}{{}^\star F}_{\nu\sigma}=G\delta^\mu_\nu$, one
determines the traceless NLE energy--momentum tensor
$\Upsilon_{\mu\nu}$  to be
\begin{eqnarray}&&\label{Upsilon}
{\Upsilon^\mu}_{\nu}:={T^\mu}_{\nu}-\frac{T}{4}{\delta^\mu}_{\nu}=
L_F({F}^{\mu\sigma}{F}_{\nu\sigma}-\,F{\delta^\mu}_{\nu}).
\end{eqnarray}
Of course, this traceless energy tensor. via the Einstein equation
is equivalent to the traceless Ricci tensor
${S^\mu}_{\nu}=\kappa{\Upsilon^\mu}_{\nu}$. Consequently it falls
into the Segre((11)(1,1))-Pleba\'nski $(2S-2T)_{(11)}$ class of
energy tensors, see Stephani et al. \cite{KramerStephani03}, Chapter
5. In the linear Maxwell limit, $ ({T^a}_b)= {\text
{diag}}(\lambda_1,\lambda_1,-\lambda_1,-\lambda_1)
 $, see  \cite{LichBook1955},pag 20.\\
Here we present the first stationary axially symmetric exact black
hole solution to the Einstein equations coupled to nonlinear
electrodynamics given by the metric
\begin{eqnarray}&&\label{metric1}
ds^2=\rho ^2\,{\bf d \theta}^2 +\,\frac{a^2\sin^2{\theta}}{\rho
^2}\left( {\bf{dt}} -\frac{a^2+r^2}{a} {\bf d \phi}
\right)^2\nonumber\\&&+\frac{\rho ^2}{Q(r)}\,{\bf
dr}^2-{\frac{Q(r)}{\rho ^2}}\left( {\bf{dt}} - a\sin^2{\theta}{\bf d
\phi} \right)^2,
\end{eqnarray}
where $\rho$ is defined as
$\rho(\theta,r):=\sqrt{{r^2+a^2\cos^2{\theta}}}$, with the
structural function $Q(r)$ of the form
\begin{eqnarray}\label{solQbeta}&&
Q(r)=\frac{\kappa\,{\it F_0}}{2}\, \left( 1-{\beta}\,{r}^{2} \right)
^{2}-2\,m\,r+{r}^{2}+{a}^{2},\nonumber\\&& {\it F_0}={\it f_1}\,{\it
y_1}-{\it g_1}\, {\it x_1},
\end{eqnarray}
where $m$ represents the mass, $a$ stands for the rotation
parameter, the constants ${f_1,g_1,y_1,x_1}$ represent the electric
and magnetic charges $e$ and $g_0$, and two parameters,${\it F_0}$
and ${\it G_0}$ related to the presence in $L$ of the invariants $F$
and $G$. Finally, the parameter $\beta$ is associated to the
nonlinearity of the electrodynamics field potentials $A_\mu$ and
${}^{\star}P_{\mu}$. In general electrodynamics $L(F,G)$, beside the
constant $F_0\neq 0$, there is a second field constant $G_0$ which
takes care of the presence of the second invariant $G $ through the
 non vanishing of $L_G$ even in the case of the linear Maxwell field.
 Without any loss of generality, one may set ${\it f1}={\it
g_0},{\it g_1}=e$ and equate
\begin{eqnarray}&&
{\it x_1}=-{\frac {{\it F_0}\,e-{\it G_0}\,{\it g_0}} {{e}^{2}+{{\it
g_0}}^{2}}},{\it y_1}={\frac {{\it F_0}\,{\it g_0}+{\it G_0}
\,e}{{e}^{2}+{{\it g_0}}^{2}}}.
\end{eqnarray}
In the Maxwell case, the Kerr--Newman solution is determined for
 $${\it f_1}={\it g_0},{\it g_1}=e,{\it y_1}={\it g_0},{\it x1}=-e,
 {\it G_0}=0,{\it F_0}={e}^{2}+{{\it g_0}}^{2},
 $$
with $L(F)= L(F_0)\neq 0$ and $L_G=0$. Therefore, as a by product,
we got the Kerr--Newman solution for a Lagrangian depending on the
two invariants $F\simeq {\bf E}^2-{\bf B}^2$ and $ G\simeq{\bf
E}\cdot{\bf B} $, although it can be determined via duality
rotations \cite{SalazarGarciaPleb1}. \\
The null tetrad $\bf{e^a}$ used is
\begin{displaymath}\label{tetradn}
 \left.\begin{array}{cc}{
 \bf {e^{1}}}\\  {\bf{e^{2}}}
\end{array}\right\}
= \frac{1}{\sqrt{2}}\,\left[\rho \,{\bf d \theta}\pm
i\,\frac{a\sin{\theta}}{\rho}\left( {\bf{dt}} -\frac{a^2+r^2}{a}
{\bf d \phi} \right)\right],
\end{displaymath}mnv
\begin{displaymath}
 \left.\begin{array}{cc}{
 \bf {e^{3}}}\\  {\bf{e^{4}}}
\end{array}\right\}
=\frac{1}{\sqrt{2}}\,\left[\frac{\sqrt{Q}}{\rho}\left( {\bf{dt}} -
a\sin^2{\theta}{\bf d \phi} \right)\pm\,\frac{\rho} {\sqrt{Q}}\,{\bf
d r}\right],
\end{displaymath}
it allows to write down the metric as
\begin{eqnarray}
g=2{\bf{e^1}\bf{e^2}}-2{\bf{e^3}\bf{e^4}}=g_{ab}{\bf{e^a}\bf{e^b}},\,\,\,
\bf{e^a}&&={e^a}_\mu \,\bf{dx}^{\mu}.
\end{eqnarray}
Additionally to the null tetrad one may introduce an orthonormal
basis $\{\bf {E^a},a=1,\ldots,4\}=\{{\bf x},{\bf y},{\bf z},{\bf
t}\}$,
 where ${\bf t}$ is a time--like vector, ${\bf t\cdot t}=-1$,
such that ${\bf e^1}=({\bf x}+i{\bf y} )/\sqrt{2}$, ${\bf e^2}=({\bf
x}-i{\bf y} )/\sqrt{2}$, ${\bf e^3}=({\bf t}+{\bf z} )/\sqrt{2}$,
${\bf e^4}=({\bf t}-{\bf z} )/\sqrt{2}$. In coordinates
$\{\theta,r,\phi,t\}$, as this basis one defines:
\begin{eqnarray}&&\label{orthob}
{\bf E^1}=\rho \,{\bf d \theta}, \, {\bf
E^3}=\frac{a\sin{\theta}}{\rho}\left( {\bf{dt}} -\frac{a^2+r^2}{a}
{\bf d \phi}\right),\,\nonumber\\&&
 {\bf E^2}=\frac{\rho}{\sqrt{Q}} \,{\bf d r},\,
 {\bf E^4}=\frac{\sqrt{Q}}{\rho}\left( {\bf{dt}} - a\sin^2{\theta}{\bf d
\phi} \right).
\end{eqnarray}
These bases are associated to the eigenvector bases; for instance;$
F_{\mu\nu}= 2F_{ab}{e^a}_{[\mu }\,{e^b}_{\nu]}=2F_{12}{e^1}_{[\mu
}\,{e^2}_{\nu]}+2F_{34}{e^3}_{[\mu }\,{e^4}_{\nu]}$ and
${\mathcal{E}^{\mu}}_{\nu}= \mathcal{E}^1_1\,{E^1}^{\mu
}\,{E^1}_{\nu }+ \mathcal{E}^2_2\,{E^2}^{\mu }\,{E^2}_{\nu }+
\mathcal{E}^3_3\,{E^3}^{\mu }\,{E^3}_{\nu }-
\mathcal{E}^4_4\,{E^4}^{\mu
}\,{E^4}_{\nu }$.\\
From the metric tensor one evaluates the coordinate components of
the Ricci tensor, the scalar curvature, and the Riemann-Weyl
curvature tensor. In particular, these curvature quantities,
refereed to the above null tetrad, acquire their simplest
description--their eigenvector structure: the Einstein tensor is
described by a diagonal matrix with two pair of eigenvectors, the
corresponding traceless Ricci tensor is described by a diagonal with
two pair of opposite in sign eigenvalues,
$\{\lambda_1,\lambda_1,-\lambda_1,-\lambda_1\}$, and hence,
according with the Segre--Pleba\'nski classification of the matter
tensors, it can only describes (linear Maxwell and nonlinear)
electrodynamics  with non--zero invariants.\\
Some researchers in vain pretend to accommodate some sort of fluids,
with one negative eigenvalue pressure in the so called Kerr--like
metric \cite{TorresFayos17}, instead of facing the hard problem of
searching solutions in the ambit they belong; the fact that these
``fluid'' models, for certain ranges of the coordinates variables,
fulfil the energy conditions, does not signify they should be {\bf
solutions} of viable matter--fields unless one abandons the
``fluid'' in favor of NLE.\\
The field tensor is endowed with four components: $ F_{\theta\phi}$,
 $F_{\theta\,t}$, $ F_{r\phi}$, $F_{rt} $; two of them are independent, and
the other two are related with them through the alignment
conditions:
 \begin{eqnarray}&& {F_{r\phi}} =-a \, \sin^{2} { \theta }
 { F_{rt}} ,\,\,{ F_{\theta\phi}}
 =-{\frac {
   {a}^{2}+{r}^{2}  }{a}}{ F_{\theta t}},
\end{eqnarray}
which allow to determine the remaining non vanishing field tensor
components ${F_{r\phi}} $ and $F_{\theta\phi}$. Since the  field
tensor ${F_{\mu\nu}}$ is a curl, it can be
 determined from its representation
$F_{\mu\nu}=A_{\nu,\mu}-A_{\mu,\nu} $. Thus, the independent
components are $F_{rt}=A_{t,r}=\frac{\partial A_t}{\partial r} $ ,
and $F_{\theta\,t}=A_{t,\theta}=\frac{\partial A_t}{\partial \theta}
$, while the dependent ones are $F_{r\phi}=A_{\phi,r}=\frac{\partial
A_\phi}{\partial r} $,$F_{\theta\phi}=A_{\phi,\theta}=\frac{\partial
A_\phi}{\partial \theta} $; the integrability of $A_\phi$ yields
$$\frac{\partial}{\partial \theta}(a  \sin^{2} { \theta }
 { F_{rt}})-\frac{\partial}{\partial r}({\frac {
   {a}^{2}+{r}^{2}  }{a}}{ F_{\theta t}})=0$$
to be integrated for $A_t$ and in turn to determine $A_\phi$.\\
The components of the vector potential field
$A_{\mu}=\{0,0,A{_\phi}, A{_t}\}$, endowed with various
electromagnetic field constants, where $\beta$ is related with the
nonlinear electromagnetic field, occur to be
\begin{eqnarray}&&\label{PotenA}
\rho^2\, A_{t}= \left( {\it f1}\,{a}^{3}  \cos ^{3} \theta
 -{\it g_1}\,{r}^{3} \right) \beta
+{{ \it f_1}\,a\cos{\theta}  +{\it g_1}\,r}, \nonumber\\&&
-\rho^2\,A_\phi=a\,\beta \left[  {a}{\it f_1}\, \left(
{a}^{2}+{r}^{2} \right) \cos^{3} {\theta} -{\it g_1}\,{r}^{3}
\sin^{2} { \theta }
 \right]
 \nonumber\\&&+
 {\it f_1}\, \left( {a}^{2}+{r}^{2} \right)\cos{\theta} + a{\it
g_1}\,r \, \sin ^{2}{\theta}.
\end{eqnarray}
The dual tensor  ${}^{\star} P_{\mu\nu}$, being a curl, is defined
in terms of the vector potential components ${}^{\star} P_\mu$ ,
similarly as the $F_{\mu\nu}$, namely $ {}^{\star}
P_{\mu\nu}={}^{\star} P_{\nu,\mu}-{}^{\star} P_{\mu,\nu}. $ These
components fulfill the alignment conditions:
\begin{equation}
{}^{\star}P_{\,r\,\phi}=-a \, \sin^{2} {\theta } \,{}^\star
P_{r\,t},\,{}^{\star}P_{\theta\,\phi\,}=-{\frac { {a}^{2}+{r}^{2}
}{a}}{}^\star P_{\theta\,t}.
\end{equation}
As independent dual components of ${}^\star P_{\mu\,\nu}$ are
considered the following: $ {}^\star
P_{\,\theta\,t}={}^{\star}P_{t\,,\theta}
,\,{{}^\star}P_{r\,t}={}^{\star}P_{t\,,r}.$ The dual vector
potential ${}^{\star} P_\mu$ is given by ${}^{\star}P_t$ and
${}^{\star}P_\phi$, which amount to
\begin{eqnarray}&&\label{PotenP}
\rho^2\,{}^{\star}P_t={\beta}({a}^{3}\,{\it x_1}\, \cos^{3} {\theta}
-{\it y_1}\,{r}^{3})+ a {\it x_1}\,\cos {\theta}+{\it y_1}\,r,
\nonumber\\&& -\rho^2\,{}^{\star}P_\phi= {a} {\beta}\left[
a\,x_1\,\left( {a}^{2}+{r}^{2} \right)\, \cos^{3}{\theta}
 \,  -  {\it y_1}{r}^{3}\, \sin^{2}
{\theta} \right] \nonumber\\&&+ {\it x_1}\,(r^2+a^2)\cos{ \theta
}+a{\it y_1}\,r\,\sin^{2} {\theta}.
\end{eqnarray}
Compare the structural similarity of $A_\mu$ and ${}^{\star}P_\mu$;
when $\beta$ is equated to zero one gets the vector potential
corresponding to the Kerr--Newman solution after identifying
properly the charges.\\
The evaluation of the curvature quantities yields:\\
the Weyl invariant $\Psi_2$ becomes
\begin{eqnarray}&&
-12\,{\left( a\cos {\theta} +ir \right) ^{3} \left( ir-a\cos \theta
\right) }\,\Psi_2(\beta)= 6\,\kappa\,{\it F_0}\nonumber\\&&-12\,m
\left( i\,a\,\cos \theta +r \right) -6\,{a}^{2} \kappa\,{\it F_0}\,{
\beta}^{2}\,{r}^{2} \cos^{2} {\theta} \nonumber\\&&+ \left( 2
\,{a}^{2} \cos^{2} {\theta} -2\,{r}^{2}-8\,i\,a\,r\,\cos {\theta}
 \right) \kappa\,{\it F_0}\,\beta,
\end{eqnarray}
while $S={S^1}_1=2\,\Phi_{(1\,1)}$ becomes
\begin{eqnarray}\label{Riccibeta}
2\,\rho^4\,S({\beta})&&=\,\kappa\,{\it F_0}\,\beta\, \left(
3\,{a}^{2}\,\beta\,{r}^{2}\,
 \cos^{2} {\theta}  \right.\nonumber\\&&\left.-{a}^{2}
  \cos^{2} {\theta}  +{r}^{2} \right) +\,\kappa\,{\it F_0},
\end{eqnarray}
finally, the scalar curvature  amounts to
\begin{eqnarray}\label{scalarCb}&&
R({\beta})=-\frac{Q^{\prime\prime}-2}{\rho^2}=2\,{\frac
{\kappa\,{\it F_0}\,\beta\, \left(1- 3\,\beta\,{r} ^{2} \right)
}{{a}^{2}  \cos^{ 2} {\theta} +{r}^{2}}}.
\end{eqnarray}
Remarkable is the simplicity and the invariant character of the
energy conditions which hold for any non linear electrodynamics
energy--momentum tensor referred to the eigenvector tetrad frame;
\begin{eqnarray}&&
\mu+p_{\theta}=\frac{2}{\kappa}\,S\geq 0,
\mu-p_{\theta}=\frac{1}{2\kappa}\,R=-\frac{1}{2}\,{T^a}_a\geq
0,\nonumber\\&& \mu+p_{r}=0,\,
 \mu=T_{\mu\nu}u^{\mu}u^{\nu}\geq 0.
\end{eqnarray}
For our solution these quantities require  the scalar curvature
(\ref{scalarCb}) ${R}(\beta)\geq 0$, and as well
(\ref{Riccibeta}), $S(\beta)={S^1}_1({\beta})\geq 0$.\\
The energy conservation ${T^{\mu\nu}}_{;\nu}=0$ is encoded in the
Einstein equations. The Einstein tensor ${E^a}_b$ possesses two
pairs of {\it different} eigenvalues, ${E^1}_1={E^3}_3$ and
${E^4}_4={E^2}_2$, given by
\begin{eqnarray}&&\label{eigenEins1}
{E^1}_1=\frac{1}{2\rho^{4}}\left( \rho^2 {Q^{\prime\prime}}
  -2\,r  {Q^{\prime}} +2\,Q -2\,{a}^{2 }
\cos^{2} {\theta} -2\,{a}^{2}\right),\nonumber\\&&
{E^4}_4=\,\left(r\,Q^{\prime}-Q+a^2-r^2\right){\rho^{-4}}.
\end{eqnarray}
With respect to the eigenvector basis, ${E^a}_{b}$ is described by a
diagonal matrix $\text{diag}({E^1}_1,{E^4}_4,{E^1}_1,{E^4}_4)$
corresponding to the energy--momentum tensor ${T^a}_{b}$ of
nonlinear electrodynamics with trace $-R=2{E^1}_1+2{E^4}_4=\kappa
T$; in the Maxwell case these eigenvalues becomes equals in pairs
but of opposite signs, due to the trace--free property. The multiple
eigenvalue $S=S^1_1$,
\begin{equation}\label{eigenEins1}
{4\rho^{4}}S= \rho^2 {Q^{\prime\prime}}
  -4\,r  {Q^{\prime}} +4\,Q -2\,{a}^{2 }
\cos^{2} {\theta} -4\,{a}^{2}+2r^2,
\end{equation}
determines the traceless Ricci tensor-matrix
 $({S^a}_b)=\text{diag}(S,-S,S,-S)$;
 its components are related with
the Einstein tensor components through $2{S^1}_1={E^1}_1-{E^4}_4$.
These eigenvalue tensor structures point on the fact that the only
(plausibly) possible field--matter tensor corresponds to the
nonlinear electrodynamics.\\
Isolating from the ${E^4}_4$--equation, $(=-\kappa\,\mu(\theta,r))$,
the first order derivative $Q^{\prime}$, and using it recursively
into the ${E^1}_1$--equation,$ (=\kappa\,p_\theta)$, one gets
\begin{equation}
\frac{{E^1}_1}{\kappa}=-\frac{\left( {r}^{2}+{a}^{2} \cos^{2}
 {\theta } \right)}{2\,r}\,  {\frac {\partial }{\partial r}}{\mu}
-{\mu} \left( \theta,r \right)=p_{\theta}.
\end{equation}
On the other hand, one of the inequalities of the weak enrgy
conditions requires ${\mu}+p_{\theta}\geq 0$, consequently
\begin{eqnarray}&&
{-\frac{1}{2\,r}\, \left( {r}^{2}+{a}^{2} \cos^{2}
 {\theta } \right) {\frac {\partial
}{\partial r}}{\mu} \geq 0} \rightarrow{{\frac {\partial }{\partial
r}}{\mu}\leq 0}.
\end{eqnarray}
Therefore, for physically reasonable NLE theories
\begin{equation}
\mu(\theta,r)=\left(\,L+ {\it L\_F} \left( {\it F_{rt }}\right)^{2}
+ \frac {{\it F_{\theta t}} {\it F_{rt}} \, {\it L\_G} }{a\,\sin
{\theta} } \right)\geq 0
\end{equation}
has to be a decreasing function in $r$ from the maximal value of the
local energy density $\mu$ at the origin.

For the solution under consideration the energy density is
\begin{equation}
{\mu} \left( \theta, r \right) =\frac{\,\kappa\,{\it
F0}}{2\rho^4}\,\left( 1+3\,\beta\,{r}^{2} \right)  \left(
1-\beta\,{r}^{2} \right).
\end{equation}
As far as to the Lagrangian function $L$ is concerned, its length
 is quite considerable. Nevertheless one can get an idea of
 its structure by calculating the
trace of the electromagnetic energy--momentum tensor, $\kappa
T^\mu_\mu= -R$, which gives the relation
\begin{equation}
L=\frac{1}{4\kappa}R+L_F \,F+ L_G\, G.
\end{equation}
One easily evaluates the quantity $L_F F+L_G G$ using the
expressions of the tensors $F_{\mu\nu}$ and ${P}_{\alpha\beta}$, or
${}^\star F_{\mu\nu}$ because of
\begin{eqnarray}&&
F^{\mu\sigma}{P}_{\sigma\nu}=F^{\mu\sigma} (L_F
\,{F}_{\sigma\nu}+L_G
\,{}^{\star}F_{\sigma\nu})\nonumber\\&&=L_F\,F^{\mu\sigma}
\,{F}_{\sigma\nu}+L_G \,F^{\mu\sigma}{}^{\star}F_{\sigma\nu},
\end{eqnarray}
which, by contracting the index $\mu$ with $\nu$, leads to
$$
F L_F + G L_G=-\frac{1}{4}\,F^{\mu\sigma}{P}_{\sigma\mu}.
$$
Since the right hand side tensors are defined  through the potential
components $A_t$, $A_\phi$, ${}^{\star}P_t$, and ${}^{\star}P_\phi$,
from (\ref{PotenA}) and (\ref{PotenP}), by differentiating them one
gets the Lagrangian $L$ in terms of the variables $\theta$ and $r$.

I am indebted to J.F. Pleba\'nski for many fruitful discussions and
recommendations in my works in nonlinear
electrodynamics.\\
This manuscript has been sent for publication to PRL in November 5,
2021,

\end{document}